\documentclass[aps,prl,twocolumn,showpacs,groupedaddress]{revtex4}
\usepackage{graphics}

\begin{document}
\title{Suppression of the vortex glass transition due to correlated defects with a persistent direction perpendicular to magnetic field}

\author{Ryusuke Ikeda and Kiyokazu Myojin}
\affiliation{
Department of Physics, Kyoto University, Kyoto 606-8502, Japan}

\date{\today}

\begin{abstract}
It is found in terms of the lowest Landau level approach for the Ginzburg-Landau model that, in bulk type II superconductors with correlated defects, such as columnar defects, with a persistent direction {\it perpendicular} to an applied field ${\bf H}$, a continuous vortex-glass transition should be depressed to a low enough temperature in the limit of weak point disorder. Based on this finding, remarkable reductions of the glass transition temperatures, seen in twin-free YBCO with columnar defects in ${\bf H} \perp c$ and {\it twinned} YBCO in ${\bf H} \parallel c$, are discussed. It is pointed out that, in both of these two situations, the critical scaling of vanishing resistivities is anisotropic in spite of an {\it isotropic} scaling of correlation lengths and hence, makes it possible to determine {\it two} critical exponents by changing the relative angle between the current and the correlated defects. 

\end{abstract}


\maketitle

It is well understood that, in three-dimensional (3D) type II superconductors with correlated line defects, a continuous vortex-glass transition (the so-called Bose-glass transition) \cite{NV} should occur in nonzero magnetic fields parallel to the line defects. Since this glass ordering is conceptually and formally equivalent to the corresponding quantum transition in 2D case, it should be described according to Ref.\cite{Fisher} 
as a long-ranged phase coherence measured by the glass correlation function 
\begin{equation}
G_{\rm G}({\bf r}-{\bf r}') = [ \, | \langle \psi({\bf r}) ( \psi({\bf r}') )^* \rangle |^2 \, ], 
\end{equation}
which was introduced to describe the vortex-glass transition in 3D due to point defects \cite{FFH}. Here, $\psi$ is the pair-field, $\langle \,\,\, \rangle$ 
denotes the thermal average, and $[\, \, \, ]$ implies the average over a quenched randomness. Examining eq.(1) under the Ginzburg-Landau (GL) hamiltonian \cite{RI6} 
\begin{eqnarray}
{\cal H}&=& \int_{\bf r} \Biggl[ \epsilon |\psi|^2 + \xi_0^2 \Biggl| \Biggl(-{\rm i}\nabla + \frac{2 \pi}{\phi_0} {\bf A} \Biggr) \psi \Biggr|^2 + \frac{b}{2} |\psi|^4 \nonumber \\ 
&+& u({\bf r})|\psi|^2 + f({\bf r}) (\nabla \times {\bf j}({\bf r}))_z 
\Biggr] 
\end{eqnarray}
with random potentials $u$ and $f$ is a natural starting point for understanding a glass transition, characterized by the disappearance of the Ohmic resistance, in vortex states as far as the transition is continuous \cite{FFH,RI6,RI1,Zla}, where ${\bf j}({\bf r}) = \xi_0^2 \psi^* (-{\rm i} \nabla + 2 \pi {\bf A}/\phi_0) \psi + {\rm c.c.}$, $b > 0$, $\epsilon \simeq (T-T_{c0})/T_{c0}$, and $\phi_0$ is the flux quantum. 

In this note, we argue that, in a field {\it perpendicular} to strong line disorder, such a {\it continuous} vortex-glass transition does not occur at nonzero temperatures as far as {\it no} point disorder is present, and hence that, with increasing point disorder, the (dimensionless) glass transition temperature $t_G= T_G/T_{c0}$ in this case is elevated in contrast to the case \cite{Fendrich} with no line disorder, where $T_{c0}$ is the zero field transition in {\it each} sample. Our interest in phenomena in this field configuration was motivated by an observation \cite{Paulius} in twin-free YBCO with heavy-ions irradiated along the $c$-axis. In contrast to the material (intrinsic) anisotropy on the width of the vortex-liquid region between ${\bf H} \parallel c$ and ${\bf H} \perp c$ cases in a unirradiated sample of cuprates, the vortex-liquid region of an irradiated sample in ${\bf H} \perp c$ is much {\it wider}, in particular in lower $H$, compared with the corresponding one in ${\bf H} \parallel c$ and has almost the same extent as that of the unirradiated sample in ${\bf H} \parallel c$. Although, at the microscopic level, a heavy-ion irradiation in cuprates may correspond to an overdoping and hence, lead to a reduction of the intrinsic material anisotropy, it is quite difficult to attribute the strange resistive broadening \cite{Paulius} ($t_G$-reduction) in ${\bf H} \perp c$ to such a change of microscopic details. Based on the present theory, such a $t_G$-reduction in ${\bf H} \perp c$ should commonly occur in the cases with weaker point defects, and a higher $t_G$ is expected in samples with a larger amount of point defects. Throughout this paper, the Bragg-glass (BrG) phase is assumed to have been destroyed or to have been pushed down to lower temperatures \cite{RI5} than $t_G$ by {\it strong} (or a high density of) {\it line} disorder. Assuming a strong line disorder is appropriate in discussing the ${\bf H} \perp c$ data in Ref.\cite{Paulius} where the ordinary Bose-glass transition, typical in cases with strong columnar defects, was seen in the irradiated case in ${\bf H} \parallel c$. In addition, the same idea will be applicable to phenomena in twinned YBCO in ${\bf H} \parallel c$, if twin boundaries are parallel in orientation to one another. A similar reduction of $t_G$ in twinned YBCO in ${\bf H} \parallel c$ \cite{Lopez} may be also a reflection of the same mechanism as above. 

\begin{figure}[t]
\scalebox{0.35}[0.35]{\includegraphics{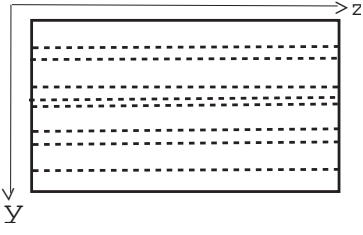}}
\caption{Superconductor with line defects (dashed lines) parallel to ${\hat z}$ described in the $y$-$z$ plane perpendicular to an applied field $\parallel {\hat x}$. The positions in the $y$-direction of defects are random.} 
\label{fig:}\end{figure}

A key fact leading to the picture suggested above is quickly found by examining the correlation function (1) within the lowest Landau level (LLL) modes of the pair-field. Since a key idea is found in the plane perpendicular to ${\bf H} \parallel {\hat x}$, as shown in Fig.1, we have only to focus on the 2D case in $y$-$z$ plane. We work in the type II limit with no gauge-fluctuation. Except in the close vicinity of $T_G$, this treatment is justified even in the case with only point disorder \cite{Kawamura}. 
Under the Landau gauge ${\bf A}= H y {\hat z}$, the pair-field within LLL is expressed by $\psi={\cal N}_0 \sum_p \varphi_p \exp({\rm i}pz - (y+p r_H^2)^2/(2 r_H^2) )$, where $r_H=\sqrt{\phi_0/(2 \pi H)}$ is the magnetic length, ${\cal N}_0$ is a normalization constant, and $p$ is a quantum number measuring the degeneracy in LLL. Then, eq.(2) in 2D takes the form ${\cal H}_0^{({\rm 2D})} + {\cal H}_{\rm pin}^{({\rm 2D})}$ within LLL, where 
\begin{eqnarray}
{\cal H}^{{(\rm 2D)}}_0 &=& (\epsilon + 2 \pi \xi_0^2 H/\phi_0) {\tilde \rho}_0 + (b/2) \sum_{\bf k} v_{k_y}  v_{k_z} |{\tilde \rho}_{\bf k}|^2, \nonumber \\
{\cal H}_{\rm pin}^{({\rm 2D})} &=& \sum_{\bf k} \Biggl( u_{\bf k} + (k_y^2 + k_z^2) r_H^2 f_{\bf k} \Biggr) (v_{k_y} v_{k_z})^{1/2} \, 
{\tilde \rho}_{-{\bf k}}. 
\end{eqnarray}
Here, $v_{k_j}=\exp(-k_j^2 r_H^2/2)$, ${\tilde \rho}_{\bf k}= \sum_p \exp({\rm i}p k_y r_H^2) (\varphi_p)^* \varphi_{p+k_z}$, and $u_{\bf k}$ ($f_{\bf k}$) is the Fourier transform of $u({\bf r})$ ($f({\bf r})$). Then, consistently, the Fourier transform of $G_{\rm G}({\bf r})$ is expressed as 
\begin{eqnarray}
{\tilde G}_{\rm G}({\bf k}) &=& v_{k_y} v_{k_z} \sum_{p,p'} e^{{\rm i}(p-p')k_y r_H^2} [ \langle \varphi_p (\varphi_{p'})^* \rangle \, \nonumber \\
&\times& \langle \varphi_{p'+k_z} (\varphi_{p+k_z})^* \rangle ]. 
\end{eqnarray}
When the random pinning potentials $u_{\bf k}$ and $f_{\bf k}$, satisfying $[u_{\bf k}]=[f_{\bf k}]=0$, are due only to line defects parallel to ${\hat z}$, i.e., persistent along ${\hat z}$, the correlators 
$[u_{\bf k} u_{{\bf k}'}]=\Delta_{\bf k} \delta_{{\bf k}+{\bf k}', \, 0}$ 
and $[f_{\bf k} f_{{\bf k}'}]=\Delta_{\Phi, {\bf k}} \delta_{{\bf k}+{\bf k}', \, 0}$ are independent of $k_z$. Since $p$ in our gauge appears as a momentum in $z$-direction, $\langle \varphi_{p_1} (\varphi_{p_2})^* \rangle$ is nonzero only when $p_1=p_2$. Then, it is clear that any $k_y$-dependence in eq.(4) occurs only from the prefactor $v_{k_y}$ which is unrelated to a critical divergence of glass correlation, and hence that $G_{\rm G}({\bf r})$ becomes short-ranged in the $y$-direction perpendicular to both the line 
defects and ${\bf H}$. It means that no 3D {\it continuous} glass transition in the present case should occur at finite temperatures ($T > 0$), because this glass correlation, according to the above-mentioned fact, is equivalent to the glass correlation in 2D systems with point disorder which does not become long-ranged at $T > 0$ \cite{Fisher,FFH,Kawamura,RI6}. 

To obtain a feeling in applying this fact to real systems, let us examine $G_{\rm G}$ in a familiar ladder approximation \cite{RI1} and using the anisotropic 3D GL model \cite{RI3} under ${\bf H} \perp c$. Since a situation with only line disorder is not possible in real materials with inevitably an amount of point disorder included, the strengths $\Delta_p$ and $\Delta_{\Phi,p}$ of point disorder and $\Delta_l$ and $\Delta_{\Phi,l}$ of line disorder are introduced in the manner $\Delta_{\bf k}= \Delta_p + \Delta_l \delta_{k_z,0}$ and $\Delta_{\Phi, {\bf k}}= \Delta_{\Phi,p} + \Delta_{\Phi,l} \delta_{k_z,0}$. We also note that, as far as the low field range \cite{RI4} $H < 0.1 \phi_0/(s^2 \gamma)$, in which the lock-in effect of Josephson vortices is negligible, is concerned, the use of the anisotropic GL model in ${\bf H} \perp c$ is justified, where $s$ is the layer spacing of a quasi-2D superconductor, and $\gamma$ is the uniaxial anisotropy. Then, after performing the random-averaging of the free energy for the random GL model, the replicated hamiltonian becomes $\sum_a {\cal H}_0^{(a)} + \sum_{a,b} {\cal H}_{\rm p}^{(a,b)}$, where $a$ and $b$ are replica indices. The replica off-diagonal term, implying the vortex pinning effects, is given by
\begin{eqnarray}
&{\cal H}&_{\rm p}^{(a,b)} \! = \! - \frac{1}{2} \int_{\bf k}  \, V(k_y, k_z) \, {\tilde \rho}^{(aa)}({\bf k}) {\tilde \rho}^{(bb)}(-{\bf k}) \nonumber \\ 
&=& \! - \int_{\bf k} \frac{{\tilde v}_{k_z}}{2} \Biggl(\Delta_p \, {\tilde v}_{k_y} + \Delta_l \frac{r_H}{\xi_0} \sqrt{\frac{2 \pi}{\gamma}} \Biggr) 
|{\tilde \rho}^{(ab)}({\bf k})|^2 ,
\end{eqnarray}
where 
\begin{equation}
V(k_y, k_z) = {\tilde v}_{k_y} \Biggl( \Delta_p {\tilde v}_{k_z} + \frac{2 \pi \Delta_l}{\xi_0} \delta(k_z) \Biggr), 
\end{equation}
${\tilde v}_{k_y}=\exp(-k_y^2 r_H^2 \gamma/2)$, ${\tilde v}_{k_z}=\exp(-k_z^2 r_H^2/(2  \gamma))$, and ${\tilde \rho}^{(ab)}({\bf k}) = \sum_p \exp({\rm i}p k_y r_H^2) (\varphi^{(a)}_p)^* \varphi^{(b)}_{p+k_z}$. For simplicity, we have dropped in eq.(5) a similar term expressed in terms of $V_\Phi(k_y, k_z)$ which is defined as $V(k_y,k_z)$ with $\Delta_p$ and $\Delta_l$ replaced by $\Delta_{\Phi,p}$ and $\Delta_{\Phi,l}$, respectively. As in the familiar case with only point disorder ($\Delta_l=\Delta_{\Phi,l}=0$), the second line of eq.(5) directly appears in the denominator of ${\tilde G}_{\rm G}({\bf k})$ in 
Gaussian approximation. In 3D, it takes the form 
\begin{equation}
{\tilde G}_{\rm G}({\bf k}) \simeq \frac{\xi_{\rm G}^2}{1 + 
\xi_{\rm G}^2 \sum_{j=x,y,z} c_j k_j^2}, 
\end{equation}
where $c_y \simeq \gamma \Delta_p$, and $c_z \simeq \gamma^{-1} (\Delta_p + \Delta_l \, r_H \sqrt{2 \pi/\gamma}/\xi_0)$. Hence, the glass correlation length $\xi_{{\rm G}, y} = c_y^{1/2} \xi_{\rm G}$ in the $y$-direction vanishes when $\Delta_p \to 0$. Physically, it means a development of an effective "anisotropy" ($\propto \Delta_p^{-1/2}$) weakening the glass correlation in the $y$-direction. Hence, a 2D-like vortex-glass fluctuation behavior should be seen when $\xi_{{\rm G}, y}$ is shorter than $r_H$. Situation is partly similar to the reduction \cite{RI4} of the rigidity of shear distortions perpendicular to the layers in the high field Josephson vortex lattice in the sense that an ordering in the direction perpendicular to both ${\bf H}$ and persistent (extended) pinning objects tends to be suppressed. Since, as well as the $\gamma$ dependence of the melting or glass transition line due only to point disorder in ${\bf H} \parallel c$, an increase of such an "anisotropy" (i.e., a reduction of correlation in a direction) reduces $T_G(H)$, the vortex liquid region should be wider as the point disorder is weakened, as far as $T_G(H)$ or $H_G(T)$ lies above, if any, a BrG melting curve (see Fig.2). This suppression of $T_G$ should be more remarkable in lower $H$ and/or in less anisotropic systems because, according to the $c_z$-expression, the effective point disorder strength to be compared with $\Delta_l$ is not $\Delta_p$ but $\Delta_p \sqrt{\gamma H}$. 

\begin{figure}[t]
\scalebox{0.25}[0.3]{\includegraphics{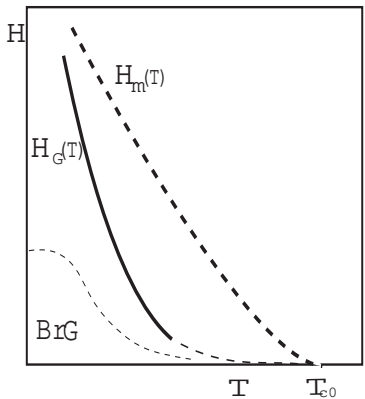}}
\caption{Vortex-glass transition (solid) curve $H_{\rm G}(T)$ in ${\bf H}$ perpendicular to strong line defects described by assuming the presence of {\it weak} point defects and a possible Bragg-glass (BrG) phase at low enough $T$. Inclusion of higher LLs will be needed to obtain the low $H$ (thin dashed) portion of $H_{\rm G}(T)$. The thick dashed curve $H_{m}(T)$ is the melting or glass transition curve in the corresponding case with no line disorder where the BrG phase may be realized just below $H_{m}(T)$ at least in lower fields. Note that $H_G(T)$ approaches $H_{m}(T)$ from {\it below} with increasing $H$. }
\label{fig:}\end{figure}

In Fig.2, an expected glass transition curve $H_G(T)$ due to line defects perpendicular to ${\bf H}$ is sketched by assuming the presence of a small amount of point disorder. As already mentioned, effects of point disorder are weaker in lower $H$ (and for less anisotropic systems), and hence, as in the figure, the solid curve $T_G(H)$ in high enough fields approaches the dashed curve from {\it lower} fields. This tendency is opposite to that in the Bose-glass case with additional point defects \cite{RI5,Klein} and seems to be consistent with features of the irreversibility lines in Ref.\cite{Paulius}. The anomalously broad vortex liquid regime in twin-free YBCO with a large amount of columnar defects \cite{Paulius} is a reflection of $t_G$ diminishing in the limit of weak point disorder, 
because the unirradiated twin-free YBCO, showing a first order melting at $H_m(T)$, should be characterized by a {\it small} amount of point disorder. 
Thus, 
any material including a larger amount of point defects should have a higher $t_G(H)$ than that seen there \cite{Paulius}. 
It is interesting to examine the corresponding situation of {\it clean} samples of more anisotropic materials, such as BSCCO and a slightly underdoped YBCO, in ${\bf H} \perp c$ which, to the best of our knowledge, have not been reported so far. 
\begin{figure}[t]
\scalebox{0.3}[0.3]{\includegraphics{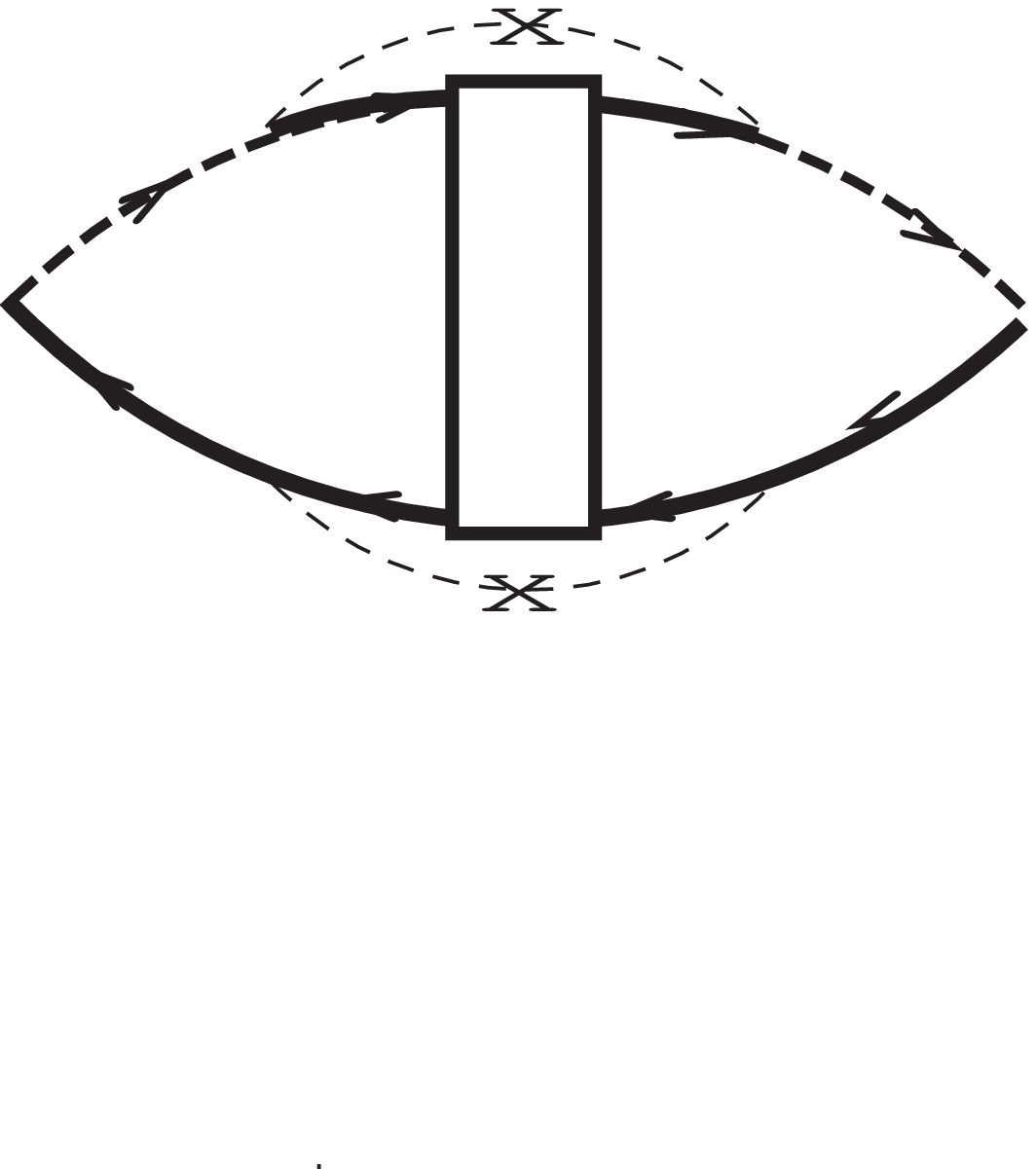}}
\caption{One of diagrams representing $\Sigma_{ll}$ ($l=y$ or $z$). The solid and thick dashed lines denote the pair-field propagators in LLL and the next lowest LL, respectively, a thin dashed line with a cross is a pinning line carrying $V(k_y,k_z)$ or $V_{\Phi}(k_y, k_z)$, and the rectangle denotes ${\tilde G}_G$.} 
\label{fig:}\end{figure}

If the line (columnar) disorder is, in contrast to the situation in Ref.\cite{Paulius}, weak enough, such an anomalously reduced $t_G(H)$ may be blocked from below by the BrG phase (see Fig.2) and not be realized, at least in low fields, as a superconducting transition line. Even in the present case with a tendency of anisotropic vortex ordering parallel to extended pinnings, it seems to us that the BrG melting should be discontinuous. \cite{com2} 

Next, let us comment on the critical behaviors of dimensionless conductivities $\Sigma_{yy}$ and $\Sigma_{zz}$ for a current perpendicular to ${\bf H} \parallel {\hat x}$ just above $T_G$. According to Refs.\cite{RI1,RI6}, a conductivity yielding the Kubo formula in a case with anisotropy in the plane perpendicular to ${\bf H}$ can be conveniently represented in terms of six pairs of Feynman diagrams composed of the glass correlation function ${\tilde G}_G({\bf k})$ and 
additional (pinning-induced) vertex corrections accompanied by $V(k_y,k_z)$ and $V_\Phi(k_y, k_z)$. One of such diagrams is described in Fig.3. 
The additional vertex correction is necessary to obtain $\Sigma_{yy}$ and $\Sigma_{zz}$ consistent with ${\tilde G}_G$ defined in LLL. The resulting glass contribution $\Sigma^{(G)}_{ll}$ to $\Sigma_{ll}$ for a current in the $l$-direction ($l=y$ or $z$) takes the form \cite{RI1}
\begin{eqnarray}
&\Sigma&^{(G)}_{ll} \! \simeq \! \int_{{\bf k}_1} \int_{{\bf k}_2} ({\bf k}_1 \times {\hat x})_l ({\bf k}_2 \times {\hat x})_l V(k_{1,y}, k_{1,z}) V_\Phi(k_{2,y}, k_{2,z}) \nonumber \\
&\times& I(k_{1,z}, k_{2,z}) \Biggl(- \frac{\partial}{\partial \Omega} \Biggr) T \sum_\omega {\tilde G}_G({\bf k}_1 - {\bf k}_2; \omega, \omega+\Omega) \Biggl|_{\Omega \to +0},
\end{eqnarray}
where $\omega$ and $\Omega$ are Matsubara frequencies, $I(k_{1,z}, k_{2,z})$ is a product of the pair-field propagators, and a positive constant factor in r.h.s. was not expressed here. The frequencies in ${\tilde G}_G$ are trivially included because the pinning functions $V$ and $V_\Phi$ carry no frequencies. If $\Delta_p=\Delta_{\Phi,p}=0$, eq.(8) results in $\Sigma^{(G)}_{yy}=0$, irrespective of the distance from 
$T_G(\Delta_p=\Delta_{\Phi,p}=0) = 0$. This should be expected because the vortices moving along the correlated defects ($ \parallel {\hat z}$) are never disturbed by them. Further, in the realistic case with point defects, eq.(8) implies 
\begin{eqnarray}
\Sigma^{(G)}_{yy} \sim \xi_G^{z_G-1}, \,\,\,\,\, 
\Sigma^{(G)}_{zz} \sim \xi_G^{z_G}, 
\end{eqnarray}
where $z_G$ is a dynamical critical exponent. That is, although the scaling of correlation lengths is isotropic, the linear response is anisotropic so that the exponent $\nu_G$ of $\xi_G$ and $z_G$ can be determined independently through resistivity data parallel {\it and} perpendicular to the line defects. Although no numerical estimation of critical exponents in this case with line defects perpendicular to ${\bf H}$ is available, it will be reasonable to expect them to take the same values as those in the purely point disorder case. 

The above-mentioned results will be applicable to other situations with correlated defects persistent in a unique direction perpendicular to ${\bf H}$. Among them, {\it twinned} YBCO samples with {\it parallel} twin boundaries (TBs) in ${\bf H} \parallel c$ are familiar and will be examined here. Since the TBs in this case have another persistent direction parallel to ${\bf H}$ leading to the transverse Meissner effect \cite{NV,RI1}, this situation has been treated so far rather as an analogue of the Bose-glass case \cite{Grigera} by neglecting a TB's persistent direction within the $a$-$b$ plane. Then, one would expect an increase of $t_G$ just like in the Bose-glass case \cite{RI5,Klein}. However, a lowering of $t_G(H)$ \cite{Lopez} due to the TBs was observed (see Fig.3 in Ref.\cite{Lopez}). As in the case with {\it both} point and line defects $\parallel {\bf H}$ \cite{RI5}, the $q_x$ dependence of $G_G({\bf q})$ peculiar to the Bose-glass case is changed near criticality, due to the point disorder in such a twinned 
sample, into the dispersion in a case with no correlated disorder. Thus, the situation is similar to the columnar-irradiated YBCO in ${\bf H} \perp c$, and the above-mentioned mechanism of a $t_G$-reduction will work well even in this twinned case although the effects may be less remarkable than in the columnar-irradiated YBCO. Actually, as seen in Fig.4 (a) of Ref.\cite{Lopez2}, an increase of point disorder in twinned YBCO seems to result in an increase of $t_G$ in contrast to the corresponding behavior in twin-free YBCO \cite{Fendrich}. 

The conductivity anisotropy in this twinned case is stronger than that of eq.(9) in the irradiated ${\bf H} \perp c$ case because of the two correlated directions of TBs, and the exponents $z_G$ and $z_G-1$ in eq.(9) are replaced in the twinned ${\bf H} \parallel c$ case by ${\tilde z}_G + 1$ and 
${\tilde z}_G - 1$, respectively, so that the ratio $\Sigma_{zz}/\Sigma_{yy} \sim \xi_G^2$ independent of the (different) dynamical exponent ${\tilde z}_G$ in this case. Here, we have assumed a different critical behavior from that in the irradiated ${\bf H} \perp c$ case. 

It is also straightforward to extend the analysis on the conductivities to static responses \cite{RI1} and to derive intuitively expected results in the twinned case such as the presence of a transverse Meissner effect for a tilt transverse to the TBs and its absence for a parallel tilt. Further, a similar anisotropy of resistive critical behavior to eq.(9) should also appear clearly in temperature dependences of resistivities in the quantum critical region of a field-tuned superconductor-insulator quantum transition \cite{Fisher,Ishida} in superconducting films with, as in Fig.1, parallel line defects in addition to point defects. In this case, $\Sigma_{zz}$ is divergent upon cooling with an algebraic power in the temperature $T$, while $\Sigma_{yy}$ saturates at a nonuniversal finite value. Their details will be explained elsewhere \cite{RI7}. 

In conclusion, a continuous vortex-glass transition due to line defects extended perpendicularly to the magnetic field is found not to occur without point defects. Based on this finding, a strange reduction of the irreversibility line in twin-free YBCO in ${\bf H} \perp c$ was discussed. In this and a similar system like a twinned YBCO in ${\bf H} \parallel c$, an {\it anisotropic} critical scaling of resistivities is realized in spite of an isotropic scaling of the correlation lengths, and its experimental verification is hoped. 

This work was supported by a Grant-in-Aid from the Ministry of Education, Culture, Sports, Science, and Technology, Japan. 


\end{document}